# Sub-milliwatt threshold power and tunable-bias all-optical nonlinear activation function using vanadium dioxide for wavelength-division multiplexing photonic neural networks

*Jorge Parra[1], Juan Navarro-Arenas[1,2], and Pablo Sanchis[1,*]*

[1]Nanophotonics Technology Center, Universitat Politècnica de València, Camino de Vera s/n, 46022, Valencia, Spain

[2]Institute of Materials Science (ICMUV), Universitat de València, Carrer del Catedràtic José Beltrán Martinez 2, 46980, Valencia, Spain

[*]Corresponding author: pabsanki@ntc.upv.es

**The increasing demand for efficient hardware in neural computation highlights the limitations of electronic-based systems in terms of speed, energy efficiency, and scalability. Wavelength-division multiplexing (WDM) photonic neural networks offer a high-bandwidth, low-latency alternative but require effective photonic activation functions. Here, we propose a power-efficient and tunable-bias all-optical nonlinear activation function using vanadium dioxide ($VO_2$) for WDM photonic neural networks. We engineered a SiN/BTO waveguide with a $VO_2$ patch to exploit the phase-change material's reversible insulator-to-metal transition (IMT) for nonlinear activation. We conducted numerical simulations to optimize the waveguide geometry and $VO_2$ parameters, minimizing propagation and coupling losses while achieving a strong nonlinear response and low-threshold activation power. Our proposed device features a sub-milliwatt threshold power, a footprint of 5 µm, and an ELU-like activation function. Temporal dynamics show a rise time as low as 5 µs. Moreover, the bias of our device could be thermally tuned, improving the speed and power efficiency. On the other hand, performance evaluations using the CIFAR-10 dataset confirmed the device's potential for convolutional neural networks (CNN). Our results show that a hybrid $VO_2$/ SiN/BTO platform could play a prominent role in the path towards the development of high-performance photonic neural networks.**

# Introduction

Artificial neural networks (ANNs), inspired by biological brains, have revolutionized computational capabilities[1], showcasing remarkable performance across a wide range of applications such as speech recognition[2], image classification[3], computer vision[4], and natural language processing[5]. Hence, ANNs have achieved human-level accuracy in tasks that are challenging for traditional computers but relatively easy for humans. In this context, the rapid advancement of ANNs has driven the need for more efficient hardware to handle their increasing complexity and computational demands[6,7]. Traditional electronic-based systems, while having powered many of these advancements, face inherent limitations in speed and energy efficiency due to the nature of electronic devices, which suffer from limited bandwidth and susceptibility to interference. As Moore's Law approaches its limits, with transistor densities nearing their physical boundaries and Dennard scaling no longer providing the expected energy efficiency gains[8], the performance growth of electronic processors has stagnated. This has led to a bottleneck in high-performance computing, where the demand for processing power continues to grow exponentially[9], particularly driven by the proliferation of artificial intelligence (AI)[10] and machine learning applications[11].

Photonic technologies present a promising solution to these limitations by leveraging light benefits[12,13]. Light offers matchless advantages in terms of bandwidth, latency, and energy efficiency. In telecommunications and data centers, optical interconnects have already established themselves as superior communication mediums due to their high-capacity and low-loss characteristics[14–16]. Therefore, this potential could be extended to information processing and computing, particularly in the domain of ANNs[17].

Integrated photonics are well-suited for high-performance implementations of ANNs due to their advantages in interconnectivity and linear operations[18,19]. Connections between pairs of artificial neurons can be represented as matrix-vector operations, with optical signals being multiplied through tunable waveguide elements and summed using wavelength-division multiplexing (WDM)[20–23]. Silicon photonics, in particular, offers a cost-effective and scalable platform, benefiting from the existing infrastructure of complementary metal oxide semiconductor (CMOS) fabrication[24–26]. However, integrated photonic neural networks still face some challenges, such as weak optical nonlinearity and the lack of suitable configurations for photonic hardware.

Strong optical nonlinearities are desired to build photonic hardware providing nonlinear activation functionalities into the neural network, thereby enabling them to model complex relationships. Current waveguide-based photonic solutions for activation functions face challenges and trade-offs related to device footprint, optical bandwidth, power threshold for activation, and dynamic tunability[27–36]. Hence, these limitations hinder the full realization of high-performance, scalable photonic neural networks. A promising approach to address these challenges could be the utilization of vanadium dioxide ($VO_2$), a phase-change material known for its reversible insulator-to-metal transition (IMT) when stimulated optically[37,38]. The IMT in $VO_2$, characterized by substantial changes in electrical conductivity and optical transmittance near room temperature (~65 ºC), presents a unique mechanism for achieving all-optical photonic devices with ultra-compact footprint and broad spectral operation such as switches[39–41], optical limiters[42], or photonic memories[43–45]. On the other hand, new CMOS-compatible integrated photonic platforms have emerged in recent years, such as the hybrid silicon or silicon nitride (SiN) – barium titanate (BTO) platform[46,47]. The SiN-BTO platform may offer significant advantages for neuromorphic computing by combining the unique properties of BTO and SiN to enable high-performance photonic integrated circuits with high operation speeds and low transmission losses[48]. In this manner, information could be transferred between electrons and photonics with ultralow heat dissipation using SiN/BTO modulators, whereas photonic information processing could be done in such $VO_2$-based devices[49].

In this work, we propose an all-optical nonlinear activation function device using $VO_2$ integrated into a SiN/BTO waveguide for WDM photonic neural networks. Our approach leverages the photoinduced IMT property of $VO_2$ to design a power-efficient and tunable bias activation function. Through numerical simulations, we optimized the waveguide geometry and $VO_2$ parameters to minimize propagation and coupling losses while achieving a strong nonlinear response and low-threshold activation power. Additionally, performance evaluations using the CIFAR-10 dataset confirm the device's potential for advanced photonic neural network applications.

## Results

**Working principle and types of $VO_2$-enabled nonlinear activation functions**

The working principle and the proposed activation function device are shown in **Fig. 1**. It consists of a SiN/BTO waveguide with a small $VO_2$ patch on top. For this work, we consider a 1.1 μm × 150 nm SiN strip resting on an 80-nm-thick BTO layer[48]. The hybrid waveguide is covered with 1 μm-thick $SiO_2$ cladding. Our proposed device works for transverse electric (TE) polarization at 1550 nm wavelength. A nonlinear response gives rise in the hybrid waveguide due to the photothermally-triggered IMT of $VO_2$ caused by evanescent coupling of the different weights applied to the input WDM signal. Hence, a fan-in operation is enabled by the non-resonant geometry of the device, together with the broad spectral response of the VO$_2$ refractive index at telecom wavelengths. On the other hand, the nonlinear input-output relationship can be engineered by tailoring the optical loss in the insulating and metallic state. The optical loss depends on the interaction between the optical mode and the VO$_2$ patch; thereby, this value can be optimized by properly designing the geometry of the hybrid waveguide and fine-tuning its parameters, such as the VO$_2$ width and thickness and the gap between the VO$_2$ patch and the SiN/BTO waveguide.

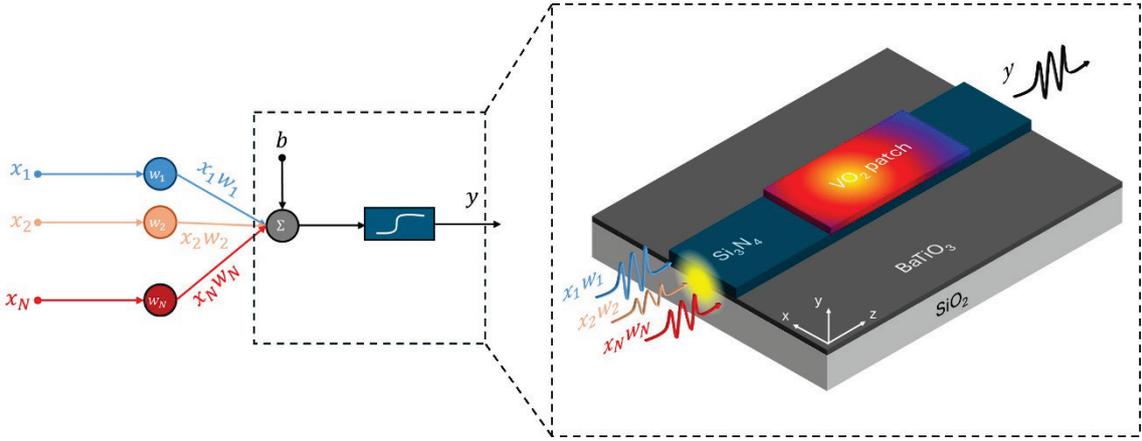

**Fig. 1.** Sketch and working principle of the proposed activation function device enabled by VO$_2$. The inputs, $x_i$, are encoded in different wavelengths and then weighted, $w_i$, accordingly by using a dot-product device before entering our device. The resulting weighted inputs are added, biased, and then evaluated by the nonlinear activation function. In our proposed device, the summation is produced by the absorption of the VO$_2$, and thus photothermally triggering its IMT, which features a nonlinear change in the refractive index of VO$_2$ with non-uniform heating in the VO$_2$ patch as a function of the optical power. Therefore, the amplitude of the output signal, $y$, will depend on the inputs in a nonlinear way. Finally, the bias of the activation function is based on the background temperature of the VO$_2$, which could be precisely controlled by using a microheater.

For this kind of waveguide, the relationship between the input and output power, in dB units, can be modeled as:

$$P_{out} = P_{in} - \int_0^L \alpha \cdot dz - 2CL \tag{1}$$

Where $P_{out}$ is the output power, $P_{in}$ is the input power, $\alpha$ is the propagation loss of the hybrid waveguide, $L$ is the length of the hybrid waveguide, and $CL$ is the coupling loss between the SiN/BTO and the VO$_2$/SiN/BTO waveguide. The parameter $\alpha$ depends on both the optical power absorbed by the VO$_2$, $P_{abs}$, which is proportional to $P_{in}$, and the position of the VO$_2$ patch along the propagation direction (z-axis), i.e., $\alpha = \alpha(P_{abs}, z)$, yielding a nonlinear input-output power relationship.

For the purpose of illustrating the potential to achieve various types of nonlinear activation functions enabled by VO$_2$, we approximate the propagation loss variation between insulating to metallic to a step-function response by considering the abrupt IMT of VO$_2$, thus:

$$\alpha \approx \begin{cases} \alpha_i, & P_{abs} < P_{IMT} \\ \alpha_m, & P_{abs} \geq P_{IMT} \end{cases} \qquad (2)$$

where $\alpha_i$ and $\alpha_m$ are the propagation loss of the optical mode in the insulating and metallic state, respectively, and $P_{IMT}$ is the optical power for which the VO$_2$ undergoes the insulator-metal change. Moreover, by considering the exponential decay of the optical power along the propagation direction, the length of VO$_2$ patch in the metallic state can be approximated as:

$$L_m \approx \begin{cases} 0, & P_{abs} < P_{IMT} \\ \eta(P_{abs} - P_{IMT}), & P_{abs} \geq P_{IMT} \end{cases} \qquad (3)$$

where $\eta$ is the thermo-optical coefficient relative to the rate conversion of the VO$_2$ length from insulator to metal[42]. Therefore, by considering negligible coupling losses, Eq. (1) simplifies as:

$$P_{out} \approx \begin{cases} P_{in} - \alpha_i L, & P_{abs} < P_{IMT} \\ P_{in} - \alpha_i L - \eta \Delta\alpha \Delta P, & P_{abs} \geq P_{IMT} \end{cases} \qquad (4)$$

where $\Delta\alpha = \alpha_m - \alpha_i$ and $\Delta P = P_{abs} - P_{th}$. The value of $\Delta\alpha$ sets the shape of the input-output relationship, thus different types of activation functions can be customized by modifying the propagation losses of the hybrid waveguide in the insulating and metallic state as shown in **Fig. 2**. If $\alpha_i > \alpha_m$ an ELU-like response is obtained [**Fig. 2(a)**], where below threshold power the output follows an exponential trend with respect to the input, while becomes linear for higher optical powers since the VO$_2$ is fully metallic. If $\eta\Delta\alpha = 1$, a flat response is achieved when the VO$_2$ patch undergoes its IMT [**Fig. 2(b)**], thus resembling to a clipped rectified linear unit (ReLU) activation function. On the other hand, if $\alpha_m > \alpha_i$ a radial basis activation function is obtained [**Fig. 2(c)**] stemming from the linear response below the threshold power and the dramatic increase of the optical loss in the metallic state.

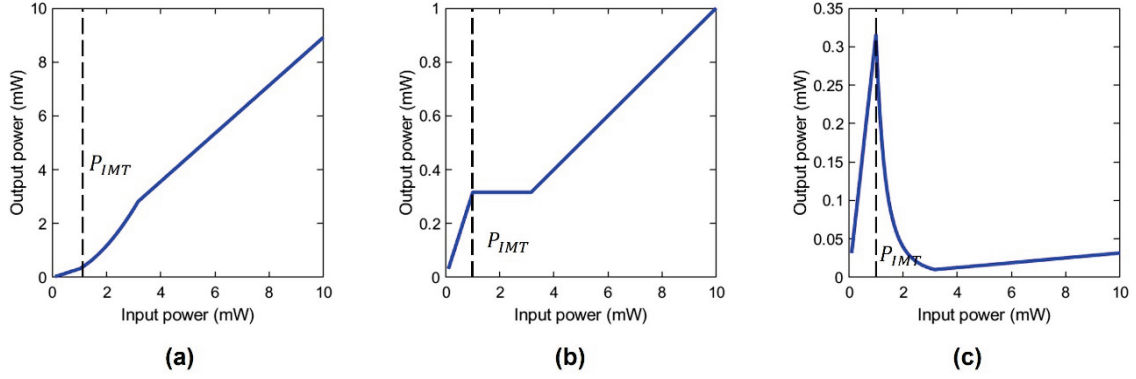

**Fig. 2.** Nonlinear activation functions based on a waveguide device and enabled by $VO_2$. **(a)** ELU. **(b)** Clipped ReLu. **(c)** Radial basis. $P_{IMT}$ stands for the threshold power of the insulator-metal transition.

**Design of the optimal $VO_2$/SiN/BTO hybrid waveguide for ELU activation function**

In this work, we target the ELU-like response to design our activation function device. In order to achieve an ELU activation function featuring a strong nonlinear response and low-threshold activation power, the cross-section of the hybrid waveguide should be optimized to achieve large propagation loss in the insulating state while minimal loss in the metallic state, along with low coupling losses in both states. To this end, we calculated the propagation losses in the insulating and metallic states as a function of the $VO_2$ thickness and gap (**Fig. 3**). We considered the $VO_2$ patch, and the SiN strip had the same width. Propagation losses were calculated from the complex effective refractive index, $n_{eff}$, of the fundamental optical mode supported by the hybrid waveguide (**Supplementary note 1**).

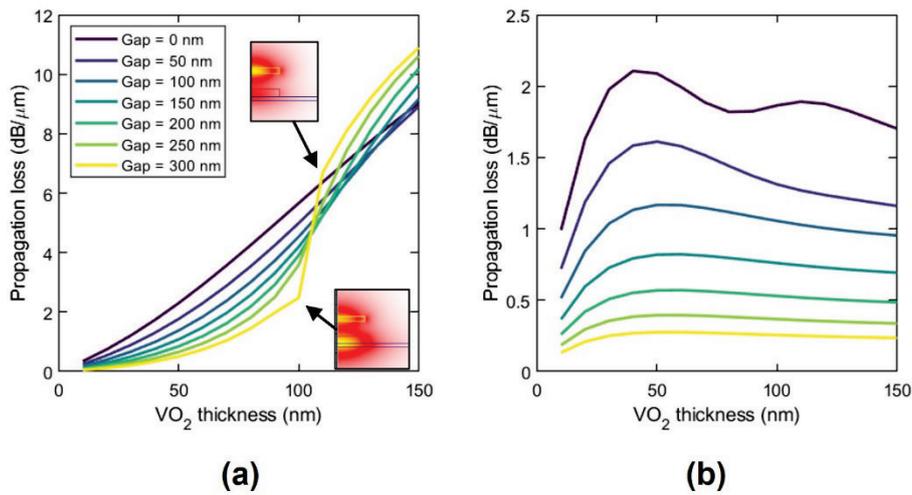

**Fig. 3.** Propagation losses as a function of the $VO_2$ thickness and gap for the **(a)** insulating and **(b)** metallic state.

In the insulating state, as the VO$_2$ becomes thicker, its interaction with the light increases since this is attracted toward the VO$_2$ due to its higher real refractive index (~2.8) compared to SiN (~2) and BTO (~2.3). Eventually, the light is highly confined inside the VO$_2$, yielding a significant increase in the propagation losses [see insets of **Fig. 3(a)**]. By contrast, in the metallic state, propagation losses show lower values and reach a plateau for thickness higher than ~50 nm, attributed to the small penetration depth of metallic VO$_2$ $\delta_p \approx \lambda/(4\pi\kappa_m) \approx 47.5$ nm. Interestingly, for small gaps, the optical mode is pushed down toward the BTO layer as the VO$_2$ thickens, resulting in a slight reduction of the propagation losses. Therefore, thick VO$_2$ layers would be desired for fulfilling the propagation loss condition ($\alpha_i > \alpha_m$) of the ELU activation function. However, large optical mismatch between the SiN/BTO and VO$_2$/SiN/BTO waveguides, and thus large coupling losses, are expected to arise for small gaps in the metallic state and for large gaps in the insulating state, thereby imposing a trade-off between low coupling losses and optimal propagation losses.

In order to evaluate the impact of the optical mismatch, we calculated the transmission of a 5-µm-long VO$_2$/SiN/BTO hybrid waveguide (**Fig. 4**) by means of 3D finite-difference time-domain (3D-FDTD) simulations (**Supplementary note 1**). On the one hand, in the insulating state and below 100 nm, the assumption of single-mode operation, along with the effective extinction coefficient provided by FEM simulations, is sufficient to predict the results of 3D-FDTD simulations; the insertion loss increases with the VO$_2$ thickness as expected [**Fig. 4(a)**]. However, for thicker VO$_2$ layers, the insertion loss of the device reduces conversely to the trend predicted by propagation losses. This effect is caused by the excitation and significant coupling of higher optical modes supported by the hybrid waveguide. These higher-order modes feature much lower propagation losses, yielding an effective propagation loss value significantly lower than predicted by considering the excitation and coupling of the fundamental mode. On the other hand, in the metallic state, we observed a deviation between FEM and 3D-FDTD loss values as a function of the VO$_2$ thickness and for gaps smaller than 150 nm [**Fig. 4(b)**]. Such a variation is due to coupling losses, which are reduced to negligible values as the VO$_2$ patch for gaps larger than 200 nm. Therefore, we choose a 100-nm-thick VO$_2$ layer with a 200 nm gap as the optimal geometry for the hybrid waveguide to build our ELU activation function device.

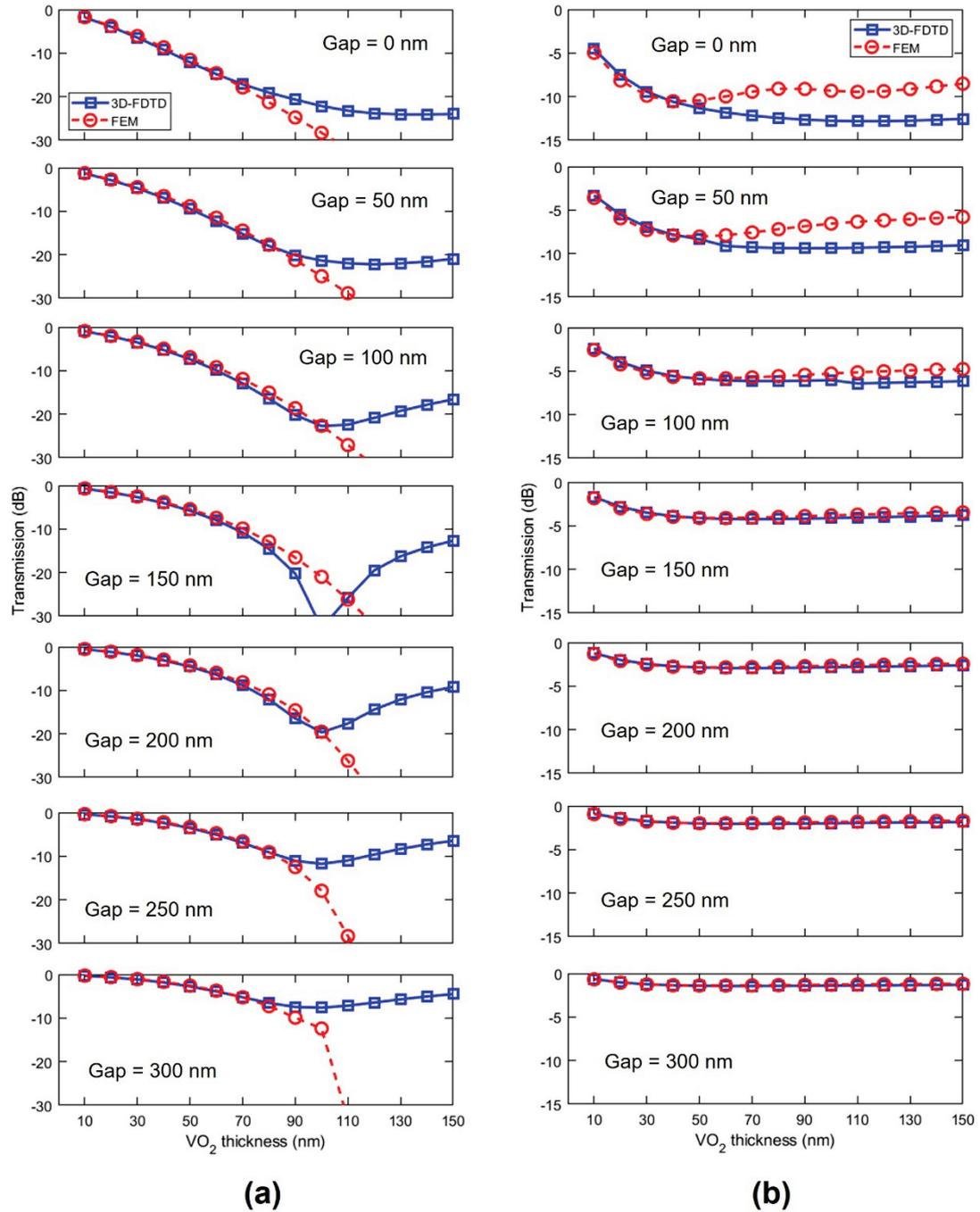

**Fig. 4.** Transmission of 5-μm-long $VO_2$/SiN/BTO waveguide as a function of the $VO_2$ thickness and gap obtained by 3D-FDTD, and comparison with calculations considering only propagation losses obtained by FEM simulations. **(a)** Insulating and **(b)** metallic states.

**Low-threshold power and bias-tunable photothermal response**

In order to determine the nonlinear optical response of our device, we carried out steady-state thermo-optical simulations to obtain the temperature distribution of the $VO_2$ patch as a function of the input power in the hybrid waveguide (**Supplementary note 2**), and thus the resulting output power.

At room temperature (20 ºC), we achieved an ELU-like response with an optical power below 0.5 mW to initiate the IMT of $VO_2$. As the input power is increased, the typical exponential response for this kind of activation function response is caused by the IMT of $VO_2$ combined with the gradual variation of the temperature along the propagation direction. On the other hand, a linear input-output relationship is obtained for optical powers higher than 2.5 mW because the $VO_2$ patch becomes fully metallic.

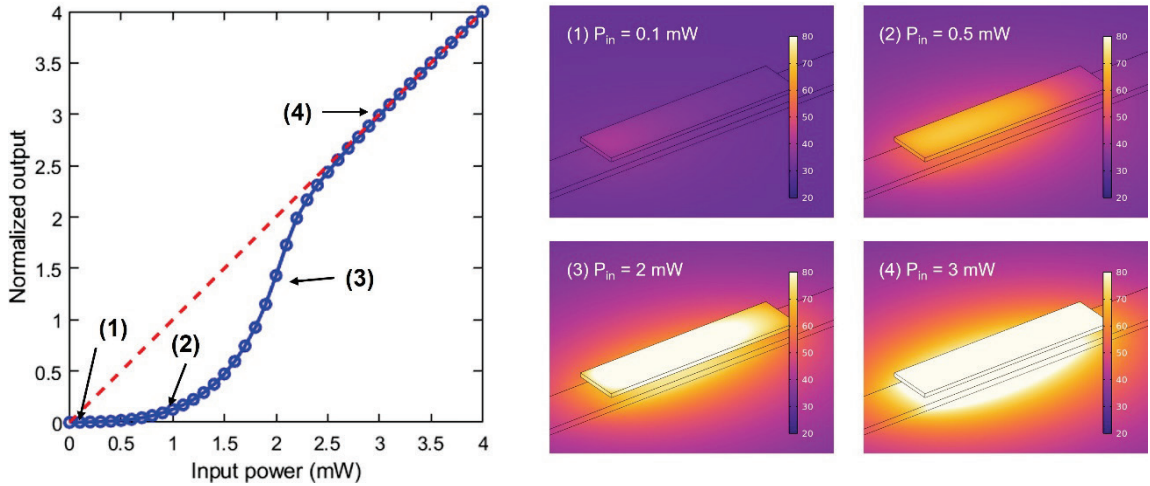

**Fig. 5.** Normalized input-output power relationship and temperature distribution (in degC) of the hybrid waveguide in the different working regimes. Results are shown for a device background temperature of 20 °C (room temperature).

We investigated the temporal dynamics of our ELU-like device by carrying out time-domain thermo-optical simulations at room temperature (**Supplementary note 2**). To this end, we modeled the input power as a rectangular pulse of 50 µs to reach the steady state of the device. We determined the rise and fall time by applying the 10-90% rule in the transmission response of the device [**Fig. 6(a)**]. For optical powers below the IMT threshold, the hybrid waveguide acts as a passive device since there is no variation of the $VO_2$ refractive index, thus, although there is a heating and cooling duration, the temporal response of the activation function is instantaneous. In its nonlinear regime, the rise time there is an interplay between the IMT and the gradual heating of the $VO_2$ patch. We observed a nonlinear relationship between the rise time and the peak power of the optical pulse [**Fig. 6(b)**]. For optical powers below 1 mW, where the output power remains almost constant with the input power (**Fig. 5**), the rise time remains around 10 µs. For higher optical power, the rise time suffers a nonlinear dependence on the input power. Finally, above 2.5 mW, the rise time is reduced down to 5 µs since the temperature increase is sufficient to uniformly change the $VO_2$ patch to the metallic state. In this case, the temporal response is mainly determined by the dimensions of the device and its thermal constants.

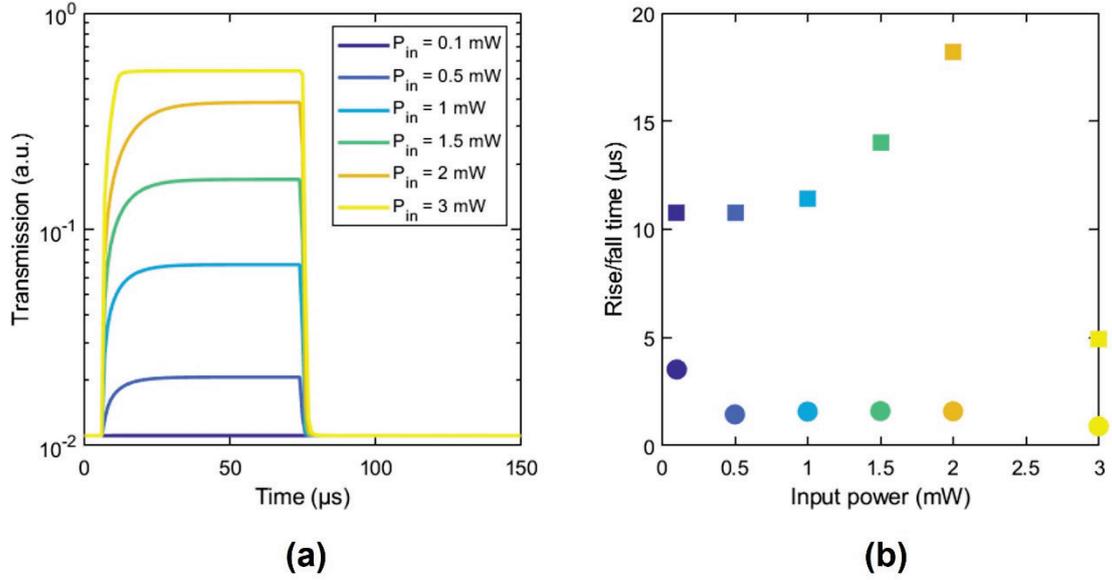

**Fig. 6. (a)** Temporal response of the activation function device when a 50-μs-wide square optical signal with different peak power, $P_{in}$, is launched into the VO$_2$/SiN/BTO waveguide. **(b)** Associated rise (squares) and fall (dots) times of the activation function as a function of input power. Results are shown for a device background temperature of 20 ºC (room temperature).

Since the working operation of the device is thermal based, the optical power for which the device's optical response changes from nonlinear to linear (threshold power, $P_{th}$) can be tuned by changing the background temperature of the device, i.e., the device response can be biased. In this regard, as the background or bias temperature is increased, the activation function is shifted toward lower input powers [**Fig. 7(a)**] because the necessary temperature increment to induce the IMT is reduced. As a consequence, the temporal response of the device is also modified [**Fig. 7(b)**]. Therefore, the threshold power and speed can be tuned as a function of the bias temperature [**Fig. 7(c)**], where sub-milliwatt threshold values and a few microsecond speeds could be achieved by setting the bias temperature in the vicinity of the IMT (~45 ºC).

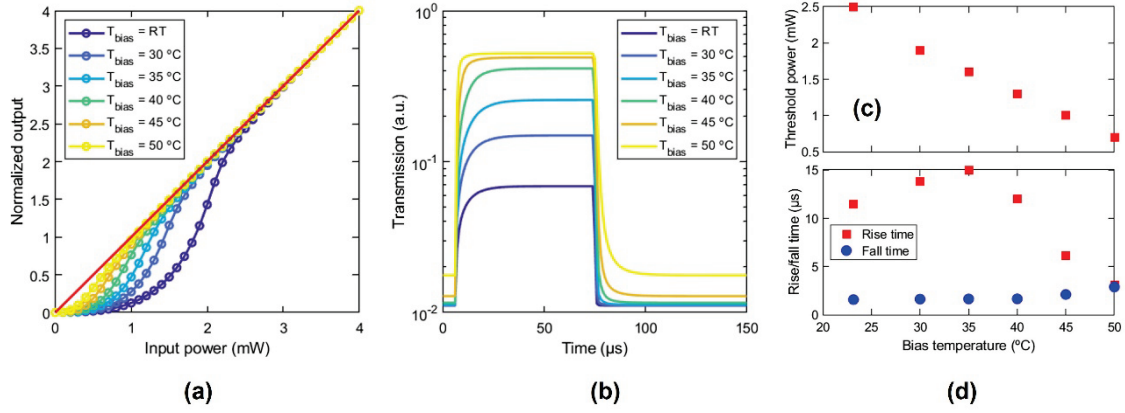

**Fig. 7.** Performance of the proposed device for different bias temperatures, $T_{bias}$. **(a)** Input-output relationship. **(b)** Temporal response under a 50-μs-wide optical pulse of 1 mW. **(c)** Threshold power and rise/fall times.

**Potential application in neural network architectures**

We evaluated the performance of the proposed hardware ELU-like activation functions in an ANN architecture. The simulated nonlinear shapes of the activation functions were employed to train a convolutional neural network (CNN) model on the CIFAR-10 dataset. In general, the ELU activation function is a well-suited choice to train CNN models. ELU offers smoothness across its entire domain, ensuring continuous gradients that facilitate more stable and efficient gradient propagation during training. Unlike ReLU, which can suffer from dead neurons (zero gradients) and saturation issues with negative inputs, ELU allows negative values, enhancing the model's ability to capture features in images.

The considered CNN model consists of two sets of convolutional layers, each followed by batch normalization and an activation function. The activation is varied to test different temperatures and compared with software ELU function implementation. These convolutional layers progressively increase in depth (32, 64, 128), allowing the model to extract hierarchical features from input images while mitigating issues like overfitting through the application of dropout layers after max-pooling operations. Max-pooling layers reduce spatial dimensions, aiding in feature extraction.

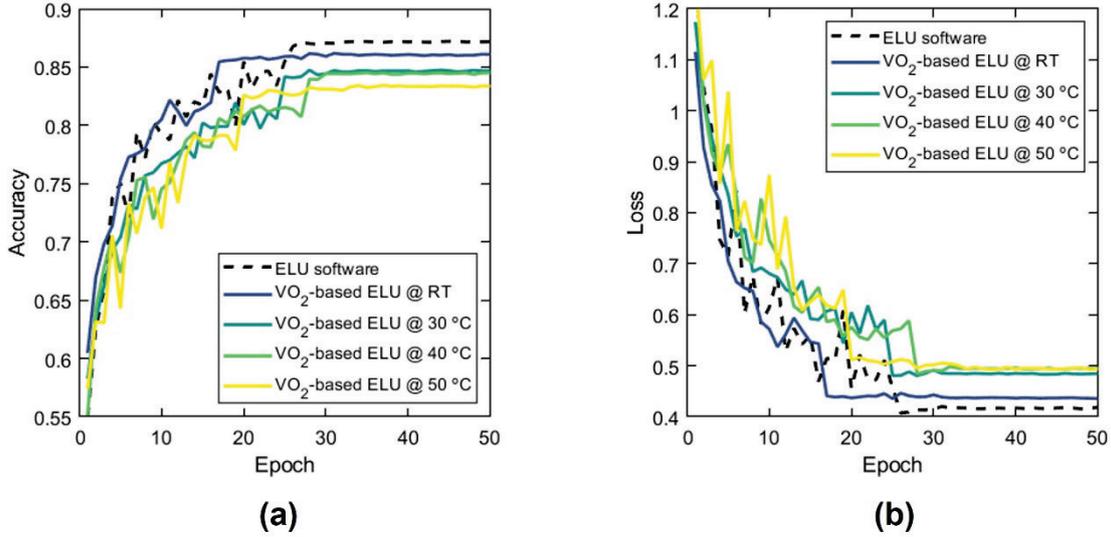

**Fig. 8**. Validation **(a)** accuracies and **(b)** losses across epochs for various activation functions, including a custom ELU function simulating the hardware response of the device at different temperatures and software ELU.

**Figure 8** shows the validation accuracy and model loss after 50 epochs of training the CNN. All functions result in high accuracy in the range of 80-85% [**Fig. 8(a)**], demonstrating that the proposed $VO_2$/Si waveguide activation function can behave similarly to an ELU function. A slight degradation of accuracy and an increase in losses are observed when the device approaches the IMT temperature [**Fig. 8(b)**]. This phenomenon can be attributed to the diminishing resemblance of the hardware response to the software ELU function (see **Fig. 7(a)**) and, thus, the loss of its inherent non-linear shape. Nonetheless, this small penalty comes with a great advantage in the threshold power and speed, as shown in **Fig. 7(d)**.

## Discussion

In this work, we have proposed an all-optical device for nonlinear activation functionalities in WDM photonic neural networks. Our device is based on the IMT of $VO_2$ integrated into a SiN/BTO waveguide and exhibits a sub-milliwatt threshold, a compact footprint of 5 $\mu m^2$, and an EKU-like nonlinear shape. The temporal dynamics of the device allow for switching speeds as low as 5 $\mu s$. On the other hand, temperature tuning enables further optimization of speed and power efficiency. Moreover, performance evaluations using the CIFAR-10 dataset confirmed the device's potential in CNN architecture.

Compared to previous integrated photonic activation function devices (**Table 1**), our proposed device would benefit from a broad spectral operation due to the non-resonant approach and the utilization of $VO_2$, which shows low wavelength dependence in their refractive index at telecom wavelengths[40,42], while maintaining a compact footprint and low-threshold power such as all-optical activation function devices based on resonant and optical bandwidth-limited structures[29–31,35]. Electro-optical activation function can benefit from using high-sensitivity photodiodes to detect low-power signals without necessitating resonant nonlinear structures[27,34–36]. However, the footprint is drastically increased considering the photodiode and associated electronic circuitry.

Therefore, our proposed device is particularly suitable for dense integrated photonics, potentially leading to significant improvements scalability and efficiency of all-optical neural networks.

**Table 1.** Survey of waveguide-based nonlinear activation function devices in integrated photonics. Nonlinearity occurs in the waveguide.

| Ref. | Simulation/ Experimental | Technology | Implemented activation functions | All-optical/O-E-O | Optical bandwidth | Threshold power | Switching speed | Footprint |
|---|---|---|---|---|---|---|---|---|
| [27] | Exp. | ITO/Si EAM | N/A | O-E-O | Broad | N/A | N/A | 5 µm |
| [28] | Exp. | Si MZI/MRR EOM | ReLU, Sigmoid | O-E-O | Limited | 0.2 mW | N/A | N/A |
| [29] | Exp. | Si MZI-MRR | Sigmoid, Clipped ReLu, Radial-basis, Softplus | All-optical | Limited | N/A | N/A | N/A |
| [30] | Exp. | Ge/Si MRR | ReLU, ELU | All-optical | Limited | 0.74 mW | 100 kHz | N/A |
| [31] | Exp. | Si MRR | Softplus, radial basis, clipped ReLU, sigmoid | All-optical | Limited | 0.08 mW | N/A | 102 µm² |
| [32] | Exp. | Ge/Si DC | Not determined | All-optical | Broad | 5.1 mW | 70 MHz | N/A |
| [33] | Exp. | Graphene/Si MRR | Not determined | O-E-O | Limited | 0.5 mW | N/A | N/A |
| [34] | Exp. | ITO/Graphene/Si EAM | Leaky ReLU | O-E-O | Broad | N/A | N/A | 1.4 µm |
| [35] | Exp. | Si MRR | Sigmoid, radial basis, negative ReLU, softplus… | O-E-O and All-optical | Limited | < 0.1 mW | GHz | N/A |
| [36] | Sim. | TCO/Si EAM | Complex | O-E-O | Broad | N/A | 100 GHz | N/A |
| This work | Sim. | VO$_2$ on SiN/BTO waveguide | Tunable ELU | All-optical | Broad | 0.5 mW | 5 µs | 5 µm² |

ITO = Indium tin oxide; Si = Silicon; EAM = Electro-absorption modulator; MZI = Mach-Zehnder interferometer; MRR = Microring resonator; Ge = Germanium; DC = Directional coupler; TCO = Transparent conductive oxide; O-E-O = Optical-electrical-optical; N/A = Not available.

## Funding

European Health and Digital Executive Agency (101070690 (PHOENIX)); Agencia Estatal de Investigación (PID2022-137787OB-I00); Generalitat Valenciana (PROMETEO Program (CIPROM/2022/14)); Ministerio de Universidades ("Margarita Salas" (MS21-037)); Universitat Politècnica de València (Grants PAID-06-23, PAID-10-23).

## Data availability

Data underlying the results presented in this paper are not publicly available at this time but may be obtained from the authors upon reasonable request.

## Competing interests

The authors declare no competing interests.

# Supplementary information

## Supplementary note 1. Optical simulations.

**Table S1** shows the refractive indices considered for optical simulations at 1550 nm wavelength.

**Table S1.** Refractive indices of the materials at 1550 nm.

| Air | BTO | SiN | SiO$_2$ | i-VO2 (T = 25 °C) | m-VO$_2$ (T = 80 °C) |
|---|---|---|---|---|---|
| 1 | 2.285 | 2.015 | 1.45 | 2.765+j0.432 | 1.789+j2.574 |

We considered the values of the VO$_2$ refractive index as a function of the temperature reported in Ref. [40]. From those values, we calculated the effective refractive index for the reported temperatures (**Fig. S1**). Then, we fitted the effective refractive indices using the Maxwell-Garnett model [**Eq. (S1)**], where the volume fraction $f$ is given by the Boltzmann function [**Eq. (S2)**].

$$\varepsilon_{EMT} = \varepsilon_i \frac{\varepsilon_m(1+2f) + \varepsilon_i(2-2f)}{\varepsilon_m(1-f) + \varepsilon_i(2+f)} \qquad (S1)$$

$$f(T) = 1 - \frac{1}{1 + \exp\left(\frac{T - T_0}{\Delta T}\right)} \qquad (S2)$$

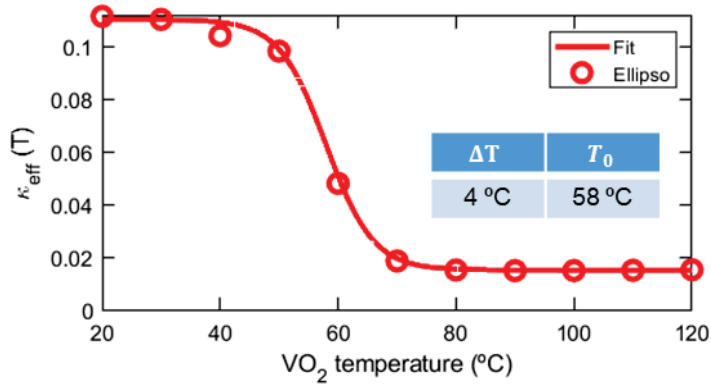

**Fig. S1.** Effective refractive index, $\kappa_{\mathrm{eff}}$, of a SiN/BTO waveguide with a 100-nm-thick VO$_2$ on top spaced by a gap of 200 nm. The table shows the fitting parameters.

We employed a finite element method (FEM) eigenmode solver (FemSIM from RSoft) to calculate the optical modes. We used a nonuniform mesh consisting of a 30 nm × 30 nm bulk grid with minimum divisions of 10 points in both the x and y axes. We used a symmetric boundary condition at x = 0 to alleviate time and computational cost. The simulated domain comprised from 0 µm to +1 µm in the x-axis, whereas from -1 µm to +1 µm in the y-axis.

The propagation loss, α, was calculated from the complex effective refractive index as:

$$\alpha = \frac{20 \log_{10}(e)\, 2\pi \kappa_{\text{eff}}}{\lambda} \tag{S3}$$

where $\kappa_{\text{eff}}$ is the imaginary part of the complex effective refractive index, and $\lambda$ is the working wavelength used in the simulation (1550 nm).

3D finite-difference time-domain (3D-FDTD) simulations using the FullWAVE tool from RSoft were conducted to calculate the insertion loss (coupling loss + propagation loss) of the device. In the XY plane, we used the same simulation configuration as for the calculation of the optical modes. In the propagation direction (z-axis), we used a 30 nm bulk grid. A perfectly matched layer (PML) was employed as a boundary condition in the remaining boundaries. Each PML boundary consisted of 10 PML cells to avoid reflections. Overlap monitors for the fundamental mode of the SiN/BTO waveguide were placed before and after the device to determine the insertion loss.

## Supplementary note 2. Thermo-optical simulations.

**Table S2** shows the thermal constants considered for thermal simulations. Thermo-optical contributions from SiN, BTO, and SiO$_2$ were considered to be negligible.

**Table S2.** Thermal constants of the materials.

|  | Si | VO$_2$ | SiO$_2$ | SiN | BTO |
|---|---|---|---|---|---|
| Thermal conductivity (W/(m K)) | 148 | 3.5 | 1.38 | 18.5 | 2.61 |
| Density (kg/m$^3$) | 2330 | 4571 | 2203 | 3100 | 5840 |
| Heat capacity (J/(kg K)) | 703 | 656 | 709 | 788 | 434 |

3D thermal simulations were conducted by solving the heat conduction equation in the steady-state and the time domain with COMSOL Multiphysics simulation tool. We modeled the VO$_2$ patch as a heat source, where heating arises from the absorption of the optical evanescent field. We approximated the heat source spatial distribution in the XY plane to be uniform, while in the z-axis (propagation direction), we considered the exponential decay, resulting thereby in the following expression:

$$Q(T,z) = \frac{P\Gamma}{wt} \frac{4\pi \kappa_{\text{eff}}(T)}{\lambda} \exp\left(-\frac{4\pi \kappa_{\text{eff}}(T)}{\lambda} z\right). \tag{S4}$$

As a consequence, the VO$_2$ patch suffers a nonlinear and nonuniform change in its refractive index profile due to the in-plane approach and the insulator-metal transition of VO$_2$.

Regarding the simulation domain and settings, the length of the SiN/BTO waveguide extended to 20 μm. The influence of the silicon substrate was taken into account, setting a size (width × height) of 20 μm × 20 μm. The height of the SiO$_2$ under- and upper-cladding was 3 μm and 1 μm, respectively. A non-uniform tetrahedral mesh was employed. The VO$_2$ patch was discretized with element sizes between 100 and 50 nm with a minimum division of 5 points in its thickness (y-axis). The SiN domain was defined by elements ranging between 100 and 200 nm. Finally, the remaining domains consisted of elements between 100 nm and 3 μm. Convective heat flux was set as the boundary condition on top of the upper cladding with a heat transfer coefficient $h = 5$ W m$^{-2}$K$^{-1}$. Temperature boundary condition was applied to the remaining boundaries with a value of 293.15 K (20 °C).

The heat conduction equation becomes highly nonlinear due to the presence of VO$_2$ as a heat source; thereby, a two-step study was defined for the stationary case. First, the temperature distribution was calculated by considering the VO$_2$ in the insulating state. Then, those results were set as initial values for solving the thermal response of the device considering its thermo-optical expression [**Eq. (S4)**]. On the other hand, for the time-dependent case, we solved the temperature distribution using **Eq. (S4)** and a time step 1/50 smaller than the width of the applied excitation pulse.